\documentclass[12pt,psfig,epsfig,graphicx,psfrag]{article}
\usepackage{dcolumn}
\usepackage{amssymb}
\usepackage{lscape}
\usepackage{amsmath}

\usepackage{amssymb}
\usepackage{amstext}
\usepackage{amsxtra}
\usepackage{mathrsfs}
\usepackage{graphicx}
\usepackage{color}
\usepackage [paper=a4paper ,lines =45 ,textheight =22cm,textwidth =15cm] {geometry}

\usepackage{t1enc}

\newcommand{\beq}{\begin{equation}}

\newcommand{\eeq}{\end{equation}}

\def\a{\alpha}
\def\b{\beta}

\def\g{\gamma}
\def\m{\mu}
\def\n{\nu}

\def\B{\Box\phi}

\newcommand{\bea}{\begin{eqnarray}}
\newcommand{\eea}{\end{eqnarray}}
\newcommand{\ba}{\begin{eqnarray}}
\newcommand{\ea}{\end{eqnarray}}

\begin{document}


\begin{center}

 {\Large \bf  k-Mouflage gravity}\\

 \vspace{0.7cm} 

\vspace{0.3cm}
 {\large E.~Babichev\footnote{babichev@apc.univ-paris7.fr},
C.~Deffayet\footnote{deffayet@iap.fr},
R.~Ziour\footnote{ziour@apc.univ-paris7.fr}}\\
{\it APC\;\footnote{UMR 7164 (CNRS, Universit\'e Paris 7 Denis Diderot, CEA, Observatoire de Paris)}, 10 rue Alice Domon et L\'eonie Duquet,\\
 75205 Paris Cedex 13, France.}\\

\vspace{0.3cm}
 \bigskip

\bigskip
\bigskip
\bigskip
\bigskip 
\bigskip

{\bf \large Abstract} 
\begin{quotation}\noindent
We introduce a large class of scalar-tensor theories where gravity becomes stronger at large distances via the exchange of a scalar that mixes with the graviton. At small distances, i.e. large curvature, the scalar is screened via an analog of the Vainshtein mechanism of massive gravity. The crossover distance between the two regimes can be made cosmological by an appropriate choice of the parameters.
\end{quotation}

\bigskip
\bigskip
\bigskip
\bigskip
\bigskip
\bigskip
\bigskip
\bigskip
\emph{Essay written for the Gravity Research Foundation 2009 Awards for\\
Essays
on Gravitation,
 awarded a honorable mention
}

\end{center}

\newpage
There has been recently a renewal of interest for theories where gravity is modified at large, cosmological, distances. This is due in part to the wish to find explanations to the observed acceleration of the expansion of the Universe not relying on a new dark component. One prototypical example of such theories is DGP gravity \cite{DGP} with its interesting cosmological consequences \cite{Deffayet:2001uy,Fifth}. Other motivations exist, like the dark matter puzzle \cite{Milgrom:1983ca} or recent observations made in Ref. \cite{Afshordi:2008rd}.
It is however notoriously difficult to modify General Relativity (GR in the following) at large distance in a consistent way without spoiling the very good agreement between GR and various experiments and observations, in particular those of the motion of solar system bodies as well as strong field tests relying on binary pulsars (see e.g. \cite{Will:2005va}). Besides DGP gravity (which suffers in particular from unsolved questions related to its UV completion), degravitation models and some related proposals \cite{DEGRAV},
some other classes of models have been considered with some details: e.g. the Chameleon models \cite{Khoury:2003aq}, or the -- equivalent to scalar-tensor --
$f(R)$ \cite{Carroll:2003wy}. What we really need is some theory that behaves as GR in the strong field limit as well as for weak fields appropriate to describe gravity in the solar system, while it deviates from GR in some ultra-weak field limit. 

A theory which was thought to have similar properties is nonlinear massive gravity, first considered in the context of strong interactions \cite{Isham:gm}. It can be defined as a bimetric theory, where one of the metrics, say $g_{\mu \nu}$, is dynamical (with an Einstein-Hilbert action) and minimally coupled to matter, while the other, say $f_{\mu \nu}$, is non dynamical and couples to $g_{\mu \nu}$ (see e.g. \cite{Damour:2002ws}). The coupling between the two metrics is chosen such that, expanding around  flat space-time, one recovers at quadratic order the Pauli-Fierz action \cite{Fierz:1939ix}, the only consistent action for a massive spin two. A massive graviton has 5 propagating polarizations, among which a scalar mode  responsible for an extra attraction felt by non relativistic sources. This scalar mode, if one only uses the quadratic Pauli-Fierz action, leads to order one difference in PPN parameters from those of General Relativity, irrespectively of the smallness of the graviton mass (this is the famous vDVZ discontinuity \cite{vanDam:1970vg}). However it was argued some time ago that the scalar polarization could hide itself in strong enough field configurations via the so-called Vainshtein mechanism \cite{Vainshtein:1972sx}. This relies on the observation that the theory has hidden derivative self interactions in the scalar sector that shuts off the effect of the scalar attraction at distances smaller than the so-called Vainshtein radius $R_V$ \cite{Deffayet:2001uk,Arkani-Hamed:2002sp}. This radius, defined by $R_V = m^{-4/5}R_S^{1/5}$ in terms of the graviton mass $m$ and the standard Schwarzschild radius of the source $R_S$, can be very large; e.g. it is much larger than the solar system  for a graviton of Hubble radius Compton length. The Vainshtein mechanism was argued to fail in nonlinear massive gravity \cite{Damour:2002gp,Jun:1986hg}, a theory which anyway suffers from ghost-like instabilities \cite{Boulware:1973my}. However, it was recently observed explicitly \cite{US}, building on some previous qualitative arguments \cite{Arkani-Hamed:2002sp,Deffayet:2005ys,Creminelli:2005qk},  that it does work in some limit of this theory where one only keeps the dominant derivative self interactions of the scalar sector, the so called "Decoupling Limit" (DL in the following).  

In this essay, we propose a general structure for a large class of scalar-tensor theories, in which the scalar field "camouflages" in strong enough gravitational fields, via a derivative self-interaction\footnote{Hence the title of this essay, where "k" refers to "kinetic".}.  Note however that to obtain this effect, the gravitational field does not need to be very strong and, similarly to massive gravity, it can already happen in the solar system.
The structure of our scalar tensor theories follows closely the one of the DL of nonlinear massive gravity and, before introducing the former, we remind some crucial properties of the latter.

 In nonlinear massive gravity, the metric field equations read 
\ba 
M_P^2 G_{\mu \nu} =\left(T_{\mu \nu}+ T^g_{\mu \nu}\right), \label{autrelab}
\ea
where $G_{\mu\nu}$ denotes
the Einstein tensor computed with the metric $g_{\mu \nu}$,
$T_{\mu \nu}$ is the matter energy momentum tensor, and
$T^g_{\mu \nu}$ is an effective energy momentum tensor coming from the interaction between metrics. It depends non derivatively on $f_{\mu \nu}$ and $g_{\mu \nu}$.
Taking a $g-$covariant derivative of Eq.(\ref{autrelab}) we obtain the constraint
\ba \label{BIAN}
\nabla^\mu T_{\mu \nu}^g =0
\ea
which $T_{\mu \nu}^g$ should obey.
Following Ref. \cite{Vainshtein:1972sx,Damour:2002gp} one can then look at spherically symmetric solutions, aimed for example at describing the metric around stars, using the ansatz
\ba
g_{\mu \nu}dx^\mu dx^\nu &=& -e^{\nu(R)} dt^2 + e^{\lambda(R)} dR^2 + R^2 d\Omega^2  \; , \label{ANG}\\
f_{\mu \nu}dx^\mu dx^\nu &=& -dt^2 + \left(1-\frac{R \mu '(R)}{2}\right)^2 e^{-\mu(R)} dR^2 + e^{-\mu(R)}R^2 d\Omega^2\; ,
\ea
where a prime denotes a derivation w.r.t. the radial coordinate $R$.
With such an ansatz, $g_{\mu \nu}$ is easy to compare with the standard form of Schwarzschild solution, while $f_{\mu \nu}$ describes a Minkowski space-time in some unsual coordinate system, parametrized by the function $\mu$ to be determined. In the DL, the system of equation to be solved, Eqs. (\ref{autrelab}-\ref{BIAN}) collapses to \cite{US}
\ba
\frac{\lambda'}{R}+\frac{\lambda}{R^{2}}&=&-\frac{1}{2}m^2 (3\mu+R\mu') + \frac{\rho}{M_P^2}, \label{E1} \\
\frac{\nu'}{R}-\frac{\lambda}{R^{2}}&=& m^2 \mu,\label{E2} \\
\frac{\lambda}{R^{2}}-\frac{\nu'}{2R}&=& Q(\mu), \label{tlambda}\\
&\equiv&  
 -
\frac{1}{2 R}\left\{3\alpha\left(6 \mu \mu'+2R \mu'^{2}+\frac{3}{2}R \mu \mu''+\frac{1}{2}R^{2} \mu ' \mu''\right) \right. \nonumber \\  
&&\left.+\beta\left(10 \mu \mu '+5R \mu'^{2}+\frac{5}{2}R \mu \mu''+\frac{3}{2}R^{2} \mu' \mu''\right)\right\},
\ea
where $\rho$ represents the source energy density, and $Q(\mu)$ contains the only left over nonlinearities parametrized by $\alpha$ and $\beta$. This parametrization covers all the possible interaction terms between the metrics \cite{US}. In the DL, those nonlinearities only appear in equation (\ref{BIAN}), which is there equivalent to Eq. (\ref{tlambda}). The first two equations (\ref{E1}-\ref{E2}) correspond to taking the DL in equation (\ref{autrelab}). $Q(\mu)$  corresponds to the strongest scalar derivative self interactions of the model \cite{US}. These interactions, when expressed in term of canonically normalized fields, are suppressed by a scale $\Lambda = m^{4/5} M_P^{1/5}$ \cite{Arkani-Hamed:2002sp} that can be made here explicit by rescaling the fields as $\nu \rightarrow M_P^{-1} \nu$, $\lambda \rightarrow M_P^{-1} \lambda$, $\mu \rightarrow M_P^{-1} m^{-2} \mu$, in which case the right hand side of equation (\ref{tlambda}) simply appears as divided by $\Lambda$. In fact the DL is simply obtained by using those rescaled fields in the original equations (\ref{autrelab}-\ref{BIAN}) and letting $M_P \rightarrow \infty$, $m\rightarrow 0$, while keeping $\Lambda$ (as well as $T_{\mu \nu}/M_P$) fixed. It is expected to give a good description of the solution in the range $R_0 \ll R \ll m^{-1}$, where $R_0$ is a distance scale that can be made parametrically lower than $R_V$, and even extend down to the Schwarzschild radius \cite{Creminelli:2005qk,US}. The structure of equations (\ref{E1}-\ref{tlambda}) can be understood from the flat space-time action 
\begin{equation}
\begin{aligned}
S =\;&\frac{M_{P}^{2}}{8}\int d^{4}x
\Big\{ 2 h^{\mu\nu} \partial_{\mu}\partial_{\nu}h - 2 h^{\mu \nu} \partial_\nu \partial_\sigma h^\sigma_\mu + h^{\mu \nu} 
 \Box h_{\mu\nu} - h \Box h\\
&\qquad+m^{2}\big[4( h_{\mu \nu} \partial^\mu \partial^\nu \phi-h \Box \phi)+ 4 \alpha \;(\Box \phi)^{3}+ 4 \beta\;(\Box \phi \;\phi_{,\mu\nu}\;\phi^{,\mu\nu})\big]\Big\}\\
&+ \frac{1}{2}\int d^{4}x\; T_{\mu\nu}h^{\mu\nu} \label{S}
\end{aligned}
\end{equation}
for the dynamical field $h_{\mu \nu}$ and $\phi$. Indeed, using the identifications \cite{US}
\ba
h_{\mu \nu} &\equiv& \{\lambda, \nu\},\label{ident1} \\
\mu &=&  -2 \phi' /R ,\label{ident2}
\ea
 $h_{\mu \nu}$ field equations correspond to Eqs. (\ref{E1}-\ref{E2}) while the one for $\phi$ correspond to Eq. (\ref{tlambda}).
The peculiarity of action (\ref{S}) is that $\phi$ does only get a kinetic term 
via a mixing with $h_{\mu \nu}$ \cite{Arkani-Hamed:2002sp}, this being entirely due to the structure of the Pauli-Fierz mass term. 
Outside of a source and at large distances, the nonlinearities $Q(\mu)$ can be neglected, and one finds that the system (\ref{E1}-\ref{tlambda}) is solved by 
\ba \label{DOMLIN}
\lambda \sim \frac{{\cal C}}{2R},\;\; \nu \sim -\frac{{\cal C}}{R},\;\; \mu \sim \frac{1}{(mR)^{2}}\frac{{\cal C}}{2R}\; .
\ea
where ${\cal C}$ is a constant of integration  expected to be proportional to $G_N$ and  to be fixed by matching to the source.
Notice that one has $\nu \sim - 2 \lambda$, in contrast with the GR result, this being due to the scalar exchange. The behaviour (\ref{DOMLIN}) is valid for $R \gg R_V$. Below $R_V$, inspection of equations (\ref{E1}-\ref{tlambda}) shows that the nonlinear term $Q(\mu)$ dominates over the terms linear in $\mu$ and at the same time is of the order of $\lambda'/R \sim \lambda/R^2\sim \nu'/R$. Hence, for $R \ll R_V$, one can neglect the terms linear in $\mu$ in front of those in the left hand side of equations (\ref{E1}) and (\ref{E2}), and as a result, one recovers linearized General Relativity (obtained for $m^2 =0$). This is the essence of the Vainshtein mechanism, and it was explictly shown to work this way by solving numerically the system of equations (\ref{E1}-\ref{tlambda}) \cite{US}. Having this in mind, one can design a large class of scalar tensor theories in which one expects to have large distance modification of GR, but a small distance recovery {\it \`a la} Vainshtein.  

Indeed, a covariantization of action (\ref{S}) is given by  
\ba
S&=&M_P^2\int d^4 x\sqrt{-g}\left(\frac{R}{2}+\frac{\g}{2}m^2\phi R + m^2 H(\phi) \right) + S_m, \label{ACTH}
\ea
where
$H(\phi)$ is some derivative (covariant and higher than quadratic) self interactions for $\phi$ that will be discussed below, and $S_m$ represents the action for matter assumed to be minimally coupled to $g_{\mu \nu}$ ($\g$ is some order one parameter). 
These theories can of course always be written in an Einstein frame. When expanded around flat space time and $\phi =0$, action (\ref{ACTH}) reproduces (\ref{S}) at quadratic order. It also contains derivative self interactions of $\phi$ similar to those left over in the DL of massive gravity and encoded into the function $H$.  
 Some possible choices for $H$ are following:
\ba
H(\phi)_{MG}&=& \frac{\a}{2}\left(\B^3\right)+\frac{\b}{2}\left(\B\,\phi_{;\m\n}\phi^{;\m\n}\right), \label{HMG}\\
H(\phi)_{DGP}&=& m^2\B\,\phi_{;\m}\phi^{;\m}, \label{HDGP}\\
H(\phi)_{K}&=& K(X), \quad {\rm with} \quad X = m^2 \phi_{;\m}\phi^{;\m}\;, \label{HK}\\
H(\phi)_{Gal} &=& m^2 \left(\phi_{;\lambda}\,\phi^{; \lambda}\right)
\left[2\left(\Box \phi\right)^2
- 2\left(\phi_{;\mu\nu}\,\phi^{;\mu\nu}\right)\right],\label{HGal} \\
H(\phi)_{CovGal} &=& m^2 \left(\phi_{;\lambda}\,\phi^{; \lambda}\right)
\left[2\left(\Box \phi\right)^2
- 2\left(\phi_{;\mu\nu}\,\phi^{;\mu\nu}\right)\label{HCovGal}
- \frac{1}{2}\left(\phi_{;\mu}\,\phi^{;\mu}\right) R\right].
\ea
The first choice is reproducing the structure of the self derivative of $\phi$ arising in the DL of massive gravity; the second choice does 
the same for DGP gravity \cite{STRONG1,STRONG1bis}\footnote{Note added after this work was
submitted to the 2009 Gravity Research Foundation Essay Competition (March
31st 2009): the recent arXiv submission \cite{Chow:2009fm} considers
aspects of the cosmology associated to a model close to (\ref{HDGP}) (see also
\cite{DPSV}).}; the third possibility, $H_K$ (chosen such that $\partial_X K(X)_{|X=0}=0$), corresponds to K-essence-like models\footnote{With however the important difference that in the Einstein frame, matter couples to a metric which depends on $\phi$.} \cite{ArmendarizPicon:1999rj}; the fourth case corresponds to the Galileon model \cite{Gal1}, while the last case is the same model with an appropriate 
non minimal coupling to curvature removing higher derivatives in the e.o.m.  \cite{Gal2} (see also \cite{Gal3}). The metric field equations obtained from (\ref{ACTH}) read 
\ba
\label{Einstein}
G_{\mu \nu} \left(1+\g m^2\phi\right)
+ \g m^2 \left(g_{\mu \nu} \B - \nabla_\mu \nabla_\nu \phi \right)
+m^2 T_{\mu \nu}^{(\phi)}
= 
M_P^{-2} \,T_{\mu \nu},
\ea
where $T_{\mu \nu}^{(\phi)}$ comes from varying $H(\phi)$. The scalar field e.o.m. is
\ba \label{SCAphi}
\frac{1}{2}\g  R + {\cal E}_\phi = 0
\ea
where ${\cal E}_\phi$ come from varying $H(\phi)$ w.r.t. $\phi$.
What is first interesting to notice is that all the models defined by (\ref{ACTH}) and their corresponding $H$ have a "Vainshtein radius" and a "decoupling limit" which isolates the strongest scalar self interaction. This interaction reads schematically 
\ba \nonumber
H(\phi)_{dom} \sim m^{2N_{f}-  N_{d}}\partial^{ N_{d}}\phi^{N_{f}},
\ea
with $N_{d}$ and $N_{f}$ integers. 
Using then the scaling (\ref{DOMLIN}) which also applies here, as well as (\ref{ident1}-\ref{ident2}),
one has
\ba \nonumber
\frac{H(\phi)}{h\B }\sim\left(\frac{R_{V,H}}{R}\right)^{N_{f}+N_{d}-4}, 
\ea
where we obtained the Vainshtein radius corresponding to the model $H$ as 
\ba \nonumber
R_{V,H}\equiv\left(R_{S}^{N_{f}-2}m^{2-N_{d}}\right)^{\frac{1}{N_{f}+N_{d}-4}}.
\ea
In the DL, and for spherical symmetry, equations (\ref{Einstein}) reduce to  
(\ref{E1}-\ref{E2}), while Eq. (\ref{SCAphi}), once integrated once, leads to an equation analogous to (\ref{tlambda})\footnote{The required integration comes from the relation (\ref{ident2}).}. The DL can be used as a starting point for a numerical integration of the full nonlinear e.o.m.  However, it is highly 
non trivial that non singular solutions found in the DL will cary into non singular solutions of the full theory, as we learned from nonlinear massive gravity \cite{Damour:2002gp,US}. 

\begin{figure}[h!]
\begin{center}$
\begin{array}{cc}
\includegraphics[width=0.47\linewidth]{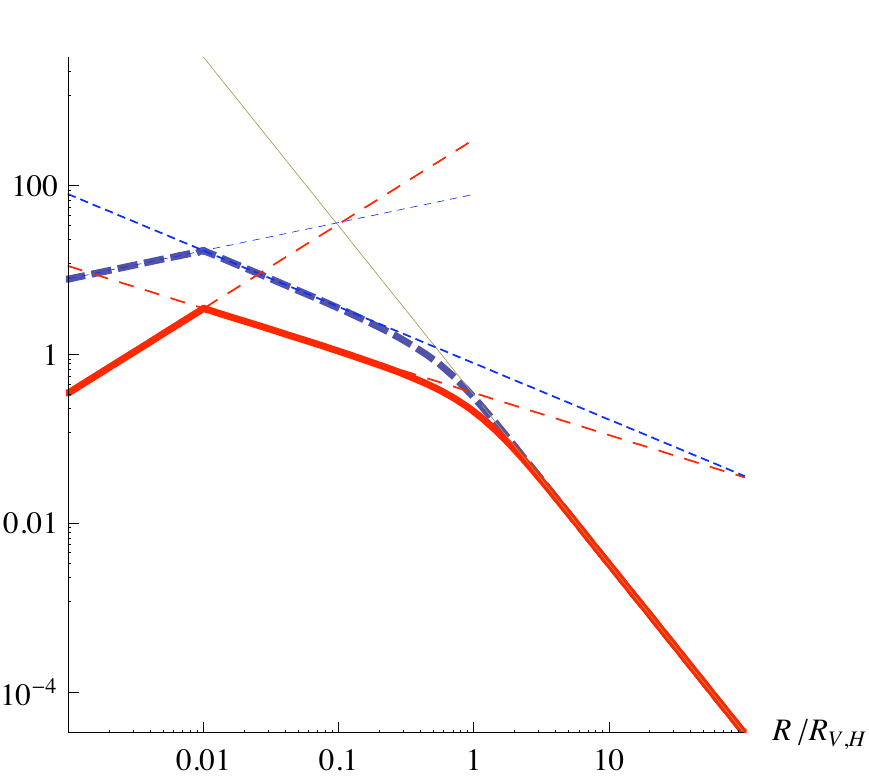} &
\includegraphics[width=0.47\linewidth]{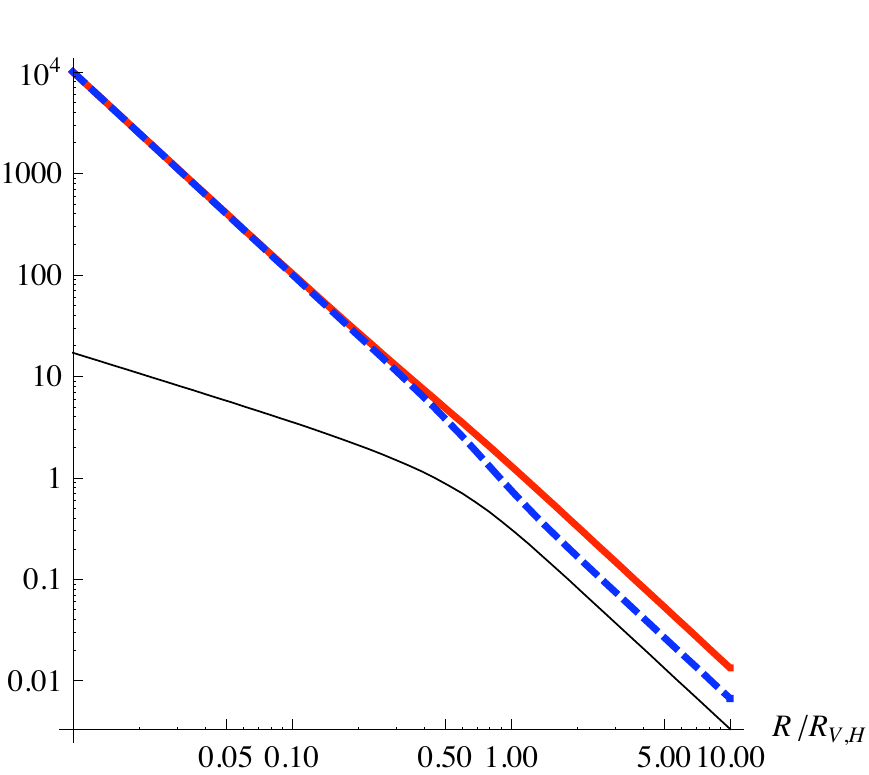}
\end{array}$
\end{center}
\caption{Left figure shows the derivative of the scalar field, $\phi'$, in the presence of a source of radius $R= 0.01$ (in the unit of the Vainshtein radii of the models) for $H \equiv H_K(X) \propto X^2$ (dashed blue curve), and $H=H_{DGP}$ (red solid line). The different asymptotic regimes are shown by thin lines. One sees in particular the transition happening at the Vainshtein radii $R_{V,H}$. The right plot shows the derivative of the functions $\lambda$ and $\nu$, $\lambda'$ (dashed blue curve) and $\nu'$ (red solid curve), appearing in the spherically symmetric ansatz (\ref{ANG}), along with $\phi'$ (thin black curve), for the $H \equiv H_K(X) \propto X^2$ model and outside the source. One sees the transition between GR regime ($R \ll R_{V,H}$) and the scalar-tensor regime ($R \gg R_{V,H}$). Below $R_{V,H}$ the scalar field "camouflages", i.e. its contribution becomes subdominant.}
\label{fig1}
\end{figure}

One of the main result reported here is that we were able to find non singular numerical solutions of the full system (\ref{Einstein}-\ref{SCAphi}) with sources, for some particular choices of $H$. Those solutions show a recovery {\it \`a la} Vainshtein of GR below the respective Vainshtein radii. Our expectation is that such solutions will also exist for a large subclass of "k-Mouflage" type of models to be determined\footnote{Note that such non singular (and in this case higher dimensional) solutions are not known explicitly in the original DGP model, where only approximate or DL solutions have been obtained  \cite{Deffayet:2001uk, Gruzinov:2001hp, STRONG1bis}.}.
Figure \ref{fig1} shows for example some of the results of the integration of the full nonlinear system for a simple $H_K$ given by $H_K(X) \propto X^2$ as well as for $H_{DGP}$. Notice in particular that the $H_{DGP}$ model has the interesting property that it has the same Vainshtein radius as the original DGP model as well as PPN corrections inside the solar system. 

Various issues are left for future works. One obvious question, not to mention possible troubles with superluminal propagations (see e.g. the different points of Refs. \cite{Adams:2006sv, Babichev:2007dw}), has to do with the stability of the different model considered. E.g. it is clear that the choices (\ref{HMG}) and (\ref{HGal}) leads to higher derivative field equations (see \cite{Gal2} for what concerns (\ref{HGal})), and hence unstable modes. Higher derivative, however, would not appear for the other choices, but this by itself does not ensure stability of the model. It is also clear that the model considered here should be UV completed in some appropriate way. The situation is the same as for DGP gravity, Galileons, or generic models of K-essence and one should worry about the low scale associated with the dominant scalar interaction. We believe however that the framework presented here, if only phenomenological, should allow various novel effects of large distance modification of gravity to be investigated in a cosmological context, a task that is certainly worth pursuing. 
\newpage
{\bf Acknowledgments}

\bigskip
 The work of E.B. was supported by the EU FP6 Marie Curie Research and Training Network UniverseNet (MRTN-CT-2006-035863). We thank Gilles Esposito-Farese for interesting discussions.

\end{document}